\documentclass{appolb}
\usepackage{epsfig}
\usepackage{amsmath,amssymb}

\newcommand{\beqn}{\begin{equation}}
\newcommand{\eeqn}{\end{equation}}
\newcommand{\req}[1]{Eq.\,(\ref{#1})}

\newcommand{\ttherm}{\tau_{\rm therm}}
\newcommand{\bag}{{\cal B}}

\begin{document}
\title{From Quark-Gluon Universe to Neutrino Decoupling: $200<T<2~{\rm MeV}$
\thanks{Presented by JR at 52 Krak\'ow School of Theoretical Physics: Astroparticle Physics in the LHC Era, Zakopane, May 19-27, 2012 }
}
\author{Michael J. Fromerth$^1$, Inga Kuznetsova$^1$,  Lance Labun$^{1,2}$,\\ Jean Letessier$^{1,3}$, and Jan Rafelski$^{1}$
\address{%
$^1$~Department of Physics, The University of Arizona,\\ Tucson, 85721 USA\\
$^2$~Leung Center for Cosmology and Particle Astrophysics, Department of Physics,\\ National Taiwan University, Taipei, Taiwan 10617\\
$^3$~Laboratoire de Physique Th{\' e}orique et Hautes Energies, Universit{\' e} Paris 6,\\ Paris 75005, France}
}
\maketitle
\begin{abstract}
The properties of the quark and hadron Universe are explored. Kinetic theory considerations are presented proving that hadron abundances after phase transformation from quarks to hadrons remain intact till abundances of hadrons become irrelevant. The hadronization process and the evolution of hadron yields is described in detail.
\end{abstract}

\section{Motivation: The big-bang and laboratory experiments}
\subsection{A short history of quark and hadron Universe}
At early times ($5<t<20~\mu{\rm s}$) in the standard model of the big bang, matter as we know it was dissolved in a state known as the quark-gluon plasma (QGP), in which leptons, gauge mesons, and quarks $u,d,s$  were deconfined and propagated freely---all hadrons are dissolved into their constituents.  At that time, strongly interacting nearly massless particles $u,d,s,G$ controlled the fate of the Universe: allowing for their interaction these  degrees of freedom comprised nearly 70\% of all energy and pressure.  The remaining 30\% was shared by $\nu_{e,\mu,\tau}, e, \mu,\gamma$~\cite{PDG12}.

The epoch we address  begins just before the quark--gluon Universe turns into the hadron Universe and ends when we reach the era of the electron, positron, neutrino, and photon  dominated Universe. Our interest is to describe in detail how hadronization in the Universe  happened and to explore the Universe's properties while it expands towards neutrino decoupling.  In outline, this epoch proceeds as follows: Around the hadronization (Hagedorn) temperature $T_h\sim 160~{\rm MeV}$, a phase transformation to the confining state occurs, and matter as we know it today forms. In the first moments this hot hadronic `gas' phase (HG) is nearly symmetric in its matter and antimatter components, and it still comprises about 70\% of all energy, so that for a short instant the Universe is hadron matter dominated.  It takes about 0.1 s for hadrons and antihadrons to annihilate so thoroughly that in essence only the net baryon number we see around us today remains: baryons in the present day Universe retain  a residual 4.6\% energy contribution, but only $10^{-10}$ fraction of particle number. Therefore, when the Universe was at a temperature prior to nucleo-syntheis but after neutrino freeze-out, let us say near $T=2$ MeV, this residual baryon matter fraction was not larger than $10^{-7}$ and the Universe was dominated by photon and neutrino `radiation'.

There is consensus that at approximately 1\,ms ($T \simeq 35$~MeV), the matter--antimatter annihilation was nearly complete. What we do not see in the literature is that the truly interesting feature of the early Universe in the following epoch is an enormous remaining pion component, with an abundance $10^{7}$ exceeding that of nucleons, and which becomes subdominant to nucleons only at $T=5$ MeV. Similarly, strangeness is present  and becomes subdominant in abundance to baryons only at $T=20$ MeV. Strangeness, and in particular CP asymmetric neutral kaons, are recreated at a time scale which is hadronic while the Universe evolves.  There are many reactions potentially enhancing the outcome of a small CP breaking in the kaon system, prompting the question, could we have too quickly discounted the possibility of baryon asymmetry originating in the hadron Universe? To answer this question we must first find an accurate and precise description of the Universe in this time period.

This report presents a brief synthesis of our effort to improve the understanding of the early quark and hadron Universe which we pursued over the past decade. Our efforts were initiated in the study of time scales of the transformation from QGP to  HG  phase~\cite{Letessier:2002gp}. We soon added to this a more detailed description of the Universe properties~\cite{Fromerth:2002wb}, assuming that chemical equilibrium prevails. Later, using kinetic theory we showed that the expansion of the Universe is indeed slow enough to assure that chemical equilibrium prevails~\cite{Kuznetsova:2008jt,Kuznetsova:2010pi}.

\subsection{Laboratory experiments addressing the early Universe}
Relativistic heavy ion collision experiments at RHIC and the LHC provide an important opportunity to simulate some aspects of these early Universe conditions.  In particular, in most central heavy ion collisions, a small nuclear-sized volume attains energy density and temperature comparable to those that prevailed in the QGP in the early Universe, this is the micro-bang.  The particles  observed later in the detectors are products of the following rapid, explosive hadronization, an energy-to-matter conversion analogous to the one occurring very slowly in the early Universe. Moreover, the QGP formed this way in the laboratory has several differences: the characteristic timescale of the heavy ion collision is much shorter $\tau\sim 10^{-23}\:{\rm s}$ which means that only strongly interacting particles equilibrate, and the baryon density is much higher, though baryon density attained at LHC  is now so low that we have yet to measure it. Taking into account these differences, we show, in this report, how the tools developed to study `micro-bangs' are applied to the early Universe hadronization and subsequent annihilation dynamics.

An important question we can address quantitatively is, what conditions in the early Universe QGP phase yield the observed matter asymmetry? The present-day small value of the baryon-to-photon ratio (which quantifies the amount of residual matter -- see section \ref{evolve_constr}) is seen by many as the result of near-complete annihilation following hadronization.  This is supported by the homogeneity of the Universe which is seen as being consistent with absence of antimatter on large scales, and some workers think  that separation into matter--antimatter domains on a scale smaller than the observable Universe is unlikely~\cite{Fixsen96,Cohen:1997ac}. 

However, more recent views on the subject sound different~\cite{Dolgov:2011zz}, we cite: ``Though observation strongly restricts possible existence of antimatter domains and objects, \ldots , it is still not excluded that antimatter may be abundant in the Universe and even in the Galaxy, not far from us. That is why there is an active search for cosmic antimatter \ldots''. A prominent example of active antimatter search is the AMS experiment~\cite{Battiston:2009zz} running at the International Space Station, searching for anti-$\alpha$. Discovery of anti-$\alpha$ would provide the first evidence that matter--antimatter separation has occurred. Such separation could arise in the early Universe in slow hadronization of QGP. We study this process  and  quantify the electrical charge distillation in Section~\ref{MixedPhase}.

We proceed starting in Section~\ref{StatMech} by first recalling the statistical physics of relativistic quantum particles, and how it is used to describe the properties and transition from the QGP to HG in the early Universe.  Special attention is given to the understanding of chemical potentials which play a pivotal role in the understanding of particle abundances: in a Universe  in thermal and chemical equilibrium  the matter--antimatter asymmetry is expressed by non-zero values of the chemical potentials. We study their magnitude quantitatively  in Section~\ref{sec:QGPHG} assuming that the Universe is on average charge neutral, the lepton number is equal to the baryon number, and entropy per baryon is derived from  baryon-to-photon ratio. 

Our study of particle abundances   relies on hadron reactions occurring even in relatively low $T$ environment at sufficient speed to assure the presence of all hadrons and we establish this considering key relaxation times in section~\ref{sec:pionlepton}. The key reaction are introduced, which connect hadrons with highly abundant photons and  leptons,   and equilibrate with the surviving pions.  The final Section~\ref{sec:concl} concludes.

\section{Statistical hadronization and the early Universe}\label{StatMech}
\subsection{Statistical hadronization, qualitatively}
A key feature of hadronic and QCD interactions is that the interactions are strong, so that at moderate temperatures or densities collisions are frequent.  In fact, they are frequent enough that statistical equilibrium is (partially) attained even within the duration of relativistic heavy ion collisions $\Delta t\sim R/c=5$-8\:fm$/c$.  The temperature and energy density achieved in the collision are sufficiently high that quarks and gluons are expected to be `deconfined' and the relevant degrees of freedom across the volume involved in the collision; in other words, a thermalized drop of QGP is created in the lab.  This dense and highly compressed state of matter expands in a `micro-bang' and as temperature drops, hadronizes. The produced yield of hadrons relates is described within a statistical hadronization approach which relies on use of phase space size to predict particle abundances.

In consequence,  relativistic heavy ion collisions evolve via  several steps, summarized here for comparison to the evolution of the early Universe plasma~\cite{Letessier:2002gp}.\\
{\bf 1.}  Initially, a small thermalized quark--gluon `fireball' is formed from the collision.  Thermalization is very rapid in the QGP; the timescale for a species $i$ is approximately its mean scattering time with all other species $j$ 
\beqn\label{ttherm}
\tau_{{\rm therm},i} \sim \frac{1}{\sum_j n_j\langle\sigma_{ij}\,v_{ij}\rangle},
\eeqn
where $\sigma_{ij}$ is the cross-section for energy-exchanging interactions, $v_{ij}\to c$ is the relative velocity of massless components and $n_j$ is number density.  For typical QGP parameters, 
\beqn
\tau_{{\rm therm},i}=0.2-2~{\rm fm}/c, \qquad 
n_j\sim 2-10~{\rm fm}^{-3},~~ \sigma_{ij}\sim 2-5~{\rm mb}.
\eeqn  
This estimate agrees with the thermalization timescale $\ttherm\lesssim 10^{-23}$~s determined from particle spectra and yields.  The microscopic mechanisms leading to such rapid thermalization are still the subject of intense study.  Even so, we expect this result to be valid qualitatively in the early Universe, and, comparing to the early Universe timescales below, implies the primordial QGP is thermalized.\\
{\bf 2.}  Chemical equilibration of quarks $u,d$ and later $s$ takes somewhat longer, $\sim 1.5$ fm$/c$ for $u,d$ and $\simeq 5$ fm$/c$ for $s$.  In general, chemical equilibration timescales are  longer than thermalization timescales, due to smaller cross sections for particle (quark pair) production as compared to energy-exchange.  However, detailed studies of the QGP properties suggest that time is sufficient for chemical equilibration of $u,d,s$ at LHC. However,  the heavy quarks $c,b$ do not have time to come to chemical equilibrium due to their higher masses implying longer chemical relaxation times. Remarkably, their abundance at in LHC heavy ion collisions is expected to be  well above chemical equilibrium due to production in initial high energy parton collisions. The same kinetic theory approach that produced relaxation times for strangeness, also allows us to compute the relaxation time of $c,b$ quarks and we find that heavy quarks are in chemical equilibrium at the time scale (microseconds) governing the end phase of QGP in  the Universe near to $T\simeq 160$\,MeV. \\
{\bf 3.}  Hadronization occurs as the fireball expands and its temperature drops. In the QGP hadronization that follows on its formation at LHC, the matter expands into empty space explosively and thus the   hadronization process is too rapid to allow re-equilibration of the final state hadrons.  When chemical equilibrium of a particular flavor can no longer be maintained in the expanding fireball, the flavor `freezes out' --- its particle number in the comoving volume remains approximately constant while its density is diluted by the expanding the volume.  Here, the early Universe differs; hadronization encompasses all the volume and  is expected to be much slower with chemical equilibrium maintained throughout.

The dynamics of chemical equilibration is especially important in relativistic plasma, in which energy can be converted to and from particles and antiparticles. Note that there are two distinct senses of chemical equilibrium:\\
{\bf i)}  {\it Absolute chemical equilibrium} is the level to which energy is shared into accessible degrees of freedom.  A black body photon gas achieves absolute chemical equilibrium almost instantaneously because the photon is massless and photons are equilibrated easily at   the hot walls of the oven.  However, a massive particle whose particle number is initially zero takes   time to achieve chemical equilibrium due to the necessity of finding particles  able to react and create the massive particle.\\
{\bf ii)}  {\it Relative chemical equilibrium} reflects the distribution of an already existent element or component  among different compounds.  Relative chemical equilibrium is thus associated with the `usual' sense of chemical potential ensuring the conservation of conserved `charges', such as baryon number.

In order to be able to describe in time the approach to absolute chemical equilibrium it is necessary to introduce, aside of chemical potentials such as $\mu_b$ for baryons, additional  abundance fugacity $\gamma_f(t)$ controlling the absolute  yields of pairs of particle $f$ so that the total fugacity, e.g., for nucleons and antinucleons is
\beqn\label{Upsilon}
\Upsilon_N=\gamma_N(t)e^{\mu_b/T},\quad 
\Upsilon_{\overline N}=\gamma_N(t)e^{\overline{\mu_b}/T}, \quad \overline{\mu_b}=-\mu_b.
\eeqn
Note that a quark is always produced paired with an antiquark so that the net baryon number is not changed.  The change in chemical potential due to an antiparticle is for this reason opposite in sign to that due to a particle as indicated above; provided, of course, the overall yield is scaled appropriately by the fugacity $\gamma(t)$ taking into account the approach to absolute chemical equilibrium.

The consequences for the energy budget are seen in the first law of thermodynamics
\begin{align}\label{1stlaw}
dE=&-P\,dV+T\,dS+T \ln \Upsilon_f\,dN+T \ln\Upsilon_{\overline{N}}\,d\overline{N},\\ 
  =&-P\,dV+T\,dS+ \mu_b (dN- d\overline{N})+T\ln\gamma_N(dN+d\overline{N}).
\end{align}
While in pair production the coefficient of $\mu_b$ vanishes,  the energy required to change the number of nucleon--antinucleon pairs is related to $\gamma_N$. Near to absolute chemical equilibrium $\gamma_N\to 1$, the last term vanishes; small fluctuations in the number of pairs do not influence the energy balance.

Statistical hadronization incorporates these dynamics of the quark and hadron `chemistry' to provide quantitative predictions of the statistical properties of the final state HG.  It successfully describes particle production in heavy ion collisions~\cite{Letessier:2002gp}, and, as a  general framework derived from the underlying theory of QCD and phenomenology of the HG phase, we can apply it to study the early Universe hadronization and the following HG dynamics.

\subsection{Statistical models of QGP and HG}
Thermodynamic properties of the QGP and HG phases are computed from the partition functions $\ln{{\cal Z}_{\rm QGP}}$ and $\ln{{\cal Z}_{\rm HG}}$; for details see~\cite{Letessier:2002gp}, we give here a short introduction.  An important difference to usual class room circumstance is that particle numbers vary. This  possibility of evolving changes in (quasi-)particle number is accounted for  in the  grand canonical partition function ${\cal Z}$
\beqn\label{Zexplicit}
{\cal Z}(\beta,V,\Upsilon_f)=V\sum_f g_f\int\!\frac{d^3p}{(2\pi)^3}\left(\ln(1\pm \Upsilon_fe^{-\beta \epsilon_f})+\ln(1\pm \Upsilon_{\bar f}e^{-\beta \epsilon_f})\right),
\eeqn
which is a function of the volume $V$, inverse temperature $\beta$ and the fugacity $\Upsilon_{f,\bar f}$ which accounts in its properties \req{Upsilon} for the possibility that absolute chemical equilibrium is not reached.  The total partition function for each component is sum over the (quasi-)particles $f$ present.  Here, $\epsilon_f^2=\vec p^2+m_f^2$ is the energy.  

For the QGP in absolute chemical equilibrium the momentum integral  can be carried out when the mass is `small' 
\beqn\begin{split}
\frac{T}{V}\ln {\cal Z}_{\rm QGP}=&-\bag+\frac{8}{45\pi^2}c_1(\pi T)^4\\
&+\sum_{f=u,d,s}\frac{1}{15\pi^2}\left(\frac{7}{4}c_2(\pi T)^4+\frac{15}{2}c_3(\mu_f^2(\pi T)^2+\frac{1}{2}\mu_f^4)\right),
\end{split}\eeqn
\beqn\nonumber
c_1=1-\frac{15\alpha_s}{4\pi}+\ldots,\quad 
c_2=1-\frac{50\alpha_s}{21\pi}+\ldots,\quad
c_3=1-\frac{2\alpha_s}{\pi}+\ldots,
\eeqn
which describes quantum gases of quarks and gluons, including also the first order perturbative QCD corrections, see the $\alpha_s$ terms in the $c_i$ coefficients, and a confining vacuum energy--pressure component $\bag \simeq 0.19$~GeV~fm$^{-3}$.  The temperature provides the scale for the evaluation of $\alpha_s(\mu)$ and is estimated as $\mu=2\pi T$ or with finite chemical potential $\mu=2\sqrt{\pi^2T^2+\mu_f^2}$.

The HG partition function, used in our computation, included a sum of partial gas contributions from all hadrons  having mass less than 2~GeV, and we apply finite volume corrections~\cite{Hagedorn:1980kb}.
Together with the QGP-liquid model~\cite{Hamieh:2000fh}, this framework is a phenomenological description of QGP equations of state which agrees  with properties of quantum chromodynamics (QCD) at finite temperature obtained in lattice QCD in the limit of vanishing particle density ($\mu_b\to 0$)~\cite{Bernard:2004je,Philipsen:2012nu,Bazavov:2012jq}.  

Recall that the statistical partition functions imply the thermodynamical relations.  Therefore, the entropy and number density in each phase can be obtained by introducing the free energy
\beqn
\frac{1}{V}{\cal F}=-\frac{T}{V}\ln {\cal Z}=-P,
\eeqn
where $P$ is the pressure. We will require below the entropy density $s$ and the baryon number density $n_B$, being 1/3 the quark density 
\beqn
s =-\frac{1}{V}\frac{d{\cal F}}{dT},\qquad
n_B =-\frac{1}{3}\frac{1}{V}\frac{d{\cal F}}{d\mu_q}.
\eeqn
See Chapters 4, 10 \& 16 of~\cite{Letessier:2002gp} for more details.

\subsection{Early Universe conditions}
First, let us check that local thermal equilibrium prevails throughout the period of interest using \req{ttherm}.  In the HG phase, where $n\sim T^{-3}$ and particle velocity is non-relativistic $v\sim T^{1/2}$, $\ttherm\sim T^{-7/2}$.   Scaling $\ttherm\sim 10^{-23}~{\rm s}$ from hadronization means  $\ttherm \lesssim 10^{-14}~{\rm s}$ at $T=1~{\rm MeV}$.  Comparing to the age of the Universe $t\sim 1~{\rm s}$ shows that the assumption of local thermal equilibrium has a large margin of error for the whole process of hadronization and equilibration discussed below.

In~\cite{Letessier:2002gp}, it is estimated  that the phase transformation from QGP to HG takes $\tau_h\sim 10\:\mu{\rm s}$ in the early Universe, many orders of magnitude longer than chemical equilibrium timescales.  Therefore, chemical equilibrium of hadrons made of $u,d,s$ quarks is firmly established at the end of the phase transformation.  Baryon annihilation reactions cease near $T=40$\,MeV.  Hadron abundance evolution in the early Universe and possible deviations from the local equilibrium will be studied down to $T\sim 2\:{\rm MeV}$ in Sec.~\ref{sec:pionlepton}. Our detailed study here demonstrates that chemical equilibrium prevails down to neutrino decoupling.  

An essential characteristic of the expansion and cooling of the Universe is that the evolution is isentropic.  Combining the first law of thermodynamics \req{1stlaw} with the Friedmann equation, there is a direct relation between the total energy density of the system and the scale factor of the Universe
\beqn\label{Friedmann}
\frac{d\epsilon}{dt}=-3\frac{1}{R}\frac{dR}{dt}(\epsilon+P),
\eeqn  
($t$ is the comoving coordinate time in Friedmann--Lema\^itre--Robertson--Walker coordinates).  For this reason, the scale factor of the Universe will remain implicit, and the system (=the Universe) will be tracked in terms of its thermodynamical properties $T,\mu_f$.  $T$ is effectively the clock tracking the progression of the system from QGP through hadronization to HG and subsequent equilibration, and, in the next section, a system of constraints relating the $\mu_f$ will be developed and solved at each $T$.  The explicit value of the expansion rate $dR/dt$ is needed only later in Sec.~\ref{sec:pionlepton}, when reaction rates must be compared to $dR/dt$ to determine freeze-out times/temperatures.

The framework so far is fixed by well-established conditions of the early Universe expansion and the statistical models of the QGP and HG phases, which are motivated by experiment.  However, we should mention several assumptions implicit in this setup and relevant to cosmology:  i) dark energy is irrelevant at this stage of the Universe's evolution; ii) dark matter does not affect the populations of visible (Standard Model) particles, neither by its decay/freeze-out nor by secondary effects; and iii) there are only left-handed neutrinos and corresponding antineutrinos.

\section{From QGP to cold hadron Gas: $200 \gtrsim T \gtrsim 2~{\rm MeV}$}
\label{sec:QGPHG}
In a system of non-interacting particles, the chemical potential $\mu_f$ of each species $f$ is independent of the chemical potentials of other species, resulting in a large number of free parameters.  In the early Universe QGP and HG, the number of independent parameters is reduced by the many chemical particle interactions occurring between the species.

First, in thermal equilibrium, photons assume the Planck distribution, implying a zero photon chemical potential: $\mu_\gamma = 0$.  Second, for any reaction $\nu_f A_f = 0$, where $\nu_f$ are the reaction equation coefficients of the chemical species $A_f$, chemical equilibrium occurs when $\nu_f \mu_f = 0$, which follows from minimizing the Gibbs free energy.

Next, chemical and thermal equilibrium means that particle--antiparticle pairs can be created from and annihilate into the thermal bath of photons, i.e., the reaction $f + \bar{f} \rightleftharpoons 2 \gamma$ proceeds freely in both directions.  Therefore, $\mu_f = -\mu_{\bar{f}}$ whenever chemical and thermal equilibrium is attained.  In particular, when the system is chemically equilibrated with respect to weak interactions, the following relationships hold, see, e.g.,~\cite{Glendenning:2000wn}:
\beqn
\mu_e - \mu_{\nu_e}=\mu_e - \mu_{\nu_{\mu}}=\mu_{\tau} - \mu_{\nu_{\tau}}\ 
\equiv\ \Delta \mu_l,
\eeqn
\beqn
\mu_u=\mu_d - \Delta \mu_l,\qquad \mu_s=\mu_d \,.
\eeqn
 The baryon chemical potential is
\beqn
\mu_b=\frac{3}{2}(\mu_d +\mu_u)=3 \mu_d -\frac{3}{2} \Delta \mu_l.
\eeqn
In relative chemical equilibrium, the chemical potential of hadrons is equal to the sum of the chemical potentials of their constituent quarks.  For example, $\Sigma^0 (uds)$ has chemical potential $\mu_{\Sigma^0}=\mu_u + \mu_d + \mu_s=3\,\mu_d - \Delta \mu_l$. 

Finally, neutrino oscillations, especially with the large mixing angle now strongly supported by experiment~\cite{An:2012eh},  imply that neutrino number is freely exchanged between flavors $\nu_e \rightleftharpoons \nu_\mu \rightleftharpoons \nu_\tau$ and hence
\beqn
\mu_{\nu_e} = \mu_{\nu_{\mu}} =\mu_{\nu_{\tau}} \equiv \mu_\nu.
\eeqn

Together, these conditions  reduce the number of independent chemical potentials to three.  We choose to track $\mu_d,\,\mu_e$, and~$\mu_\nu$.  
\subsection{Three constraints}\label{evolve_constr}
When the Universe is in a single phase (QGP or HG phase), the three free chemical potentials are determined by the following three criteria:
\begin{itemize}
\item[i.] {\it Charge neutrality} ($Q = 0$) is required to eliminate Coulomb energy.  This is written as
\beqn\label{Q0}
n_Q\equiv \sum_f\, Q_f\, n_f (\mu_f, T)=0, 
\eeqn
where $Q_i$ is the charge of species $f$, and the sum is over all particle species present in the considered particle phase.
\item[ii.] {\it Net lepton number equals net baryon number} ($L = B$) is phenomenologically motivated in the context of baryogenesis.  This leads to the constraint
\begin{equation}\label{LB0}
n_L - n_B\equiv \sum_f\, (L_f - B_f)\, n_f (\mu_f, T)=0 ,
\end{equation}
where $L_f$ and $B_f$ are the lepton and baryon numbers of species $f$. This condition is of course not proved by experiment and future studies must consider drastically different variants.
\item[iii.] {\it Constant entropy-per-baryon} ($S/B$) is equivalent to the statement that the Universe evolves adiabatically, and hence can be written
\begin{equation}
\frac{s}{n_B}\equiv\frac{\sum_f\, s_f(\mu_f, T)}{\sum_f\, B_f\, n_f(\mu_f, T)}
 = ~{\rm const},
\label{SperB}
\end{equation}
where $s_f$ is the entropy density of species $f$.
\end{itemize}

The constant $s/n_B$ is estimated from the value of the ratio
\beqn\label{eta}
\eta\equiv \frac{n_\gamma}{n_B},\quad \eta_{10}\equiv 10^{10}\eta.
\eeqn

\begin{figure}
\centerline{\includegraphics[width=0.65\textwidth]{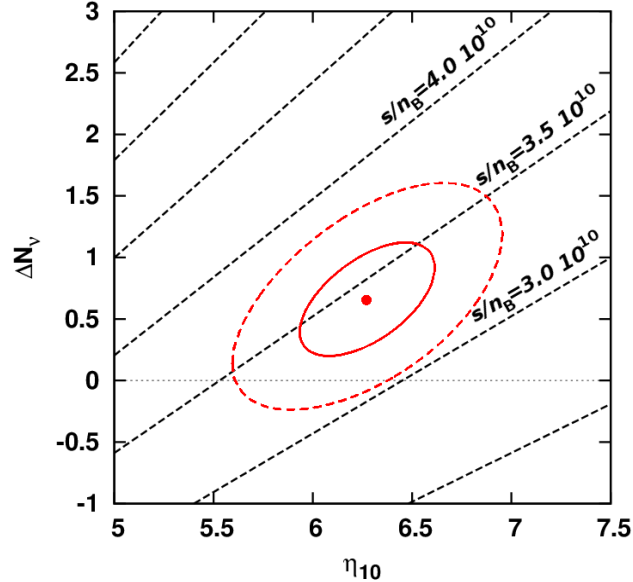}}
\caption{The fit of $\eta$ and $\Delta N_\nu$ from BBN (deuterium and $^4$He abundances) obtained in Figure 4 of~\cite{Steigman:2012ve}.  The major axes of the 68\% (solid) and 95\% (dashed) confidence ellipses align with the overlaid contours of constant $s/n_B$.  Each contour means a change of $s/n_B$ by $0.5\:10^{10}$.  \label{SperBcontours} }
\end{figure}

At $100\ge T\ge 2~{\rm MeV}$, there are 43/4 degrees of freedom~\cite{PDG12} due to  photons, $e,\mu,\tau$-neutrinos, and electron-positron pairs, which provide the dominant fraction of the entropy and  pressure, controlling the dynamics of the Universe and thus the speed of expansion at the time   big bang nucleosynthesis (BBN) occurs. We are interested in the Universe entropy content per baryon
\beqn\label{SperB1}
\frac{s}{n_B}=\frac{1}{n_B}\sum_{i=\gamma,\nu,e}s_i
=\left(\frac{s_{\gamma}}{n_{\gamma}}+\frac{n_\nu}{n_\gamma}\frac{s_\nu}{n_\nu}+\frac{n_e}{n_\gamma}\frac{s_e}{n_e}\right)\frac{n_\gamma}{n_B}, 
\eeqn
In the Standard Model, with only left-handed neutrinos, the relative number of degrees of freedom of each species is known and relates the number densities by $(4/3)n_e=2n_{\gamma}$ and $(4/3)n_{\nu}=3n_{\gamma}$.  The factor $4/3$ is the bose to fermi ratio of particle densities arising from the quantum statistics of the massless quanta.  The entropy per particle for a boson is $(s/n)_{\rm boson}=3.601$ and for a fermion is $(s/n)_{\rm fermion}=4.202$.  Inserting this into \req{SperB1} we find the known outcome. 

However, recent BBN analysis combined with  the cosmic microwave background (CMB) fluctuations have suggested that the number of relativistic degrees of freedom present at $T>1~{\rm MeV}$ is somewhat higher than the Standard Model expectation~\cite{Steigman:2012ve}.  This possibility is discussed in terms of extra neutrino degrees of freedom $3_\nu\to 3_\nu+\Delta N_\nu$, the $\Delta N_\nu$--$\eta$ fits from BBN, see circles (red) showing the best fits in Fig.~\ref{SperBcontours},  taken from  Figure 4 of~\cite{Steigman:2012ve}. The BBN best fit values~\cite{Steigman:2012ve} arise for $\Delta N_\nu=0.66^{+0.47}_{-0.45}$ and for $\eta=n_B/n_\gamma=6.27\pm0.34\:10^{-10}$. 

We   generalize \req{SperB1}  to variable neutrino abundance and obtain 
\begin{align}\label{SperBcalc}
\frac{s}{n_B}&   
 =\left(3.601 
+(3+\Delta N_\nu)\frac{3}{4} 4.202 
+2\frac{3}{4} 4.202 
\right) \frac{n_\gamma}{n_B},\\[0.2cm]
&=(19.35+3.15\Delta N_\nu)\frac{n_\gamma}{n_B}.
\end{align}
Note that this relation applies at/after BBN. We assume that there is no change in entropy content thus 
$s/n_B$ we find at time of BBN applies in the Universe prior to BBN. Baryon number is conserved.

We are interested to understand what this means for the variable $s/n_B$ which enters our evaluation of the Universe properties. In Fig.~\ref{SperBcontours}, we overlay contours of constant $s/n_B$ on Figure 4 of~\cite{Steigman:2012ve} with lines of constant  $s/n_B$. We recognize that the relevant thermodynamic variable in these considerations is the entropy per baryon, and this analysis leads to $s/n_B=3.42^{+0.42}_{-0.39}\:10^{10} $. In the following figures, however, the numerical results shown  are based on an older value obtained for $s/n_B=4.5^{+1.4}_{-1.1}\:10^{10}$ based on similar analysis of 10 years ago~\cite{Steigman:2002qv}.  Since this value of $s/n_B$ is consistent within its $1\sigma$ expectation with the updated 2012 value, we can expect the quantitative outcome changes only slightly.
 
\subsection{Chemical potentials and particle abundances in early Universe}
\begin{figure}
\centerline{\includegraphics[width=0.95\textwidth]{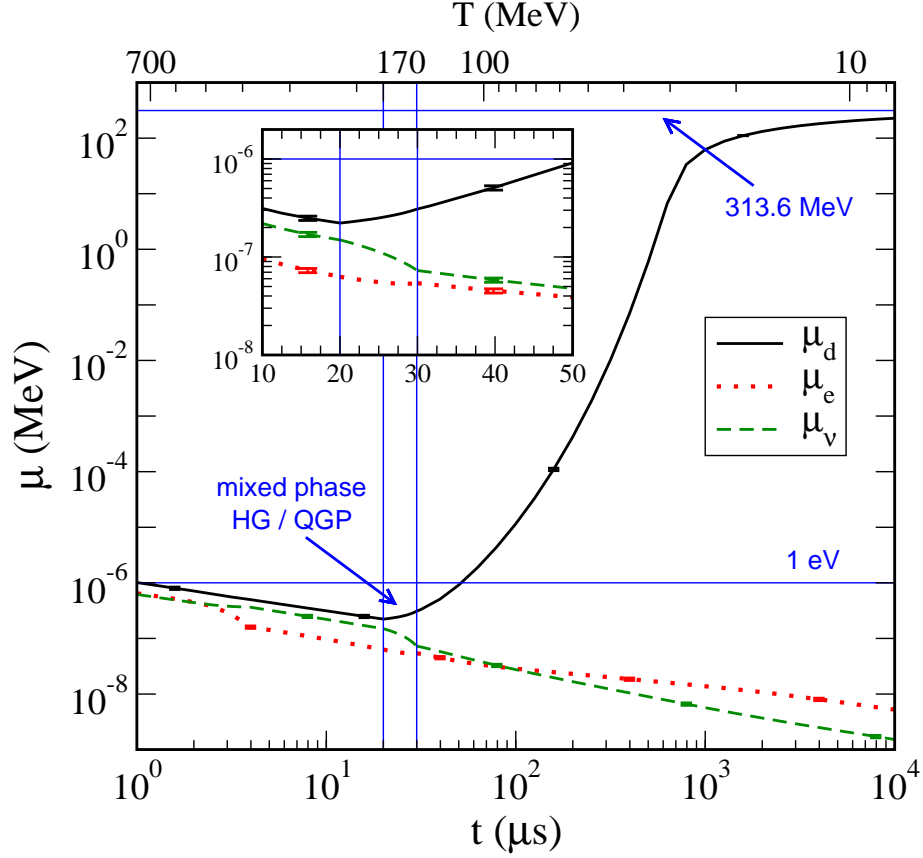}}
\caption{Chemical potentials $\mu_d$, $\mu_e$, and $\mu_\nu$ tracked from the QGP through the phase transformation to the HG.  The error bars correspond to the uncertainty in $\eta$, see below \req{SperB}.  At right, the time immediately around the transformation is zoomed.\label{chem_pot} }
\end{figure}

For each temperature $T$, the conditions Eqs.~(\ref{Q0})--(\ref{SperB}) form a system of three coupled, nonlinear equations of the three chosen unknowns ($\mu_d$, $\mu_e$, and $\mu_\nu$).  These equations were solved numerically~\cite{Fromerth:2002wb} using the Levenberg--Marquardt method~\cite{Press:1992zz} and the results shown in Fig.~\ref{chem_pot}.  The bottom axis shows the age of the Universe and the top axis the corresponding temperature.  At low temperature, $\mu_d$ approaches (weighted) one-third the nucleon mass $(2m_n-m_p)/3=313.6~{\rm MeV}$ reflecting the dominance of protons and neutrons in their classical Boltzmann limit $T\ll m_f$.

From Fig.~\ref{chem_pot}, the value of the baryon chemical potential just before the phase transition is
\beqn\label{preHGmuB}
\mu_B = 0.33^{+0.11}_{-0.08}~{\rm eV}.
\eeqn
More generally, we can say with near certainty that the values of the chemical potentials required to generate the observed matter--antimatter asymmetry are
\beqn\label{preHGchempots}
\mu_f\lesssim 0.5~{\rm eV}, \quad f=d,e,\nu\,.
\eeqn
This statement admits a generous margin of error for uncertainty in the transition temperature $T_h\sim 160~{\rm MeV}$ as well as the displayed error bars, which show the effect of the uncertainty in $s/n_B$ obtained from $\eta$.  These error bars could be improved using the updated measurements of $\eta$~\cite{Komatsu:2010fb}.  

\begin{figure}[tbp]
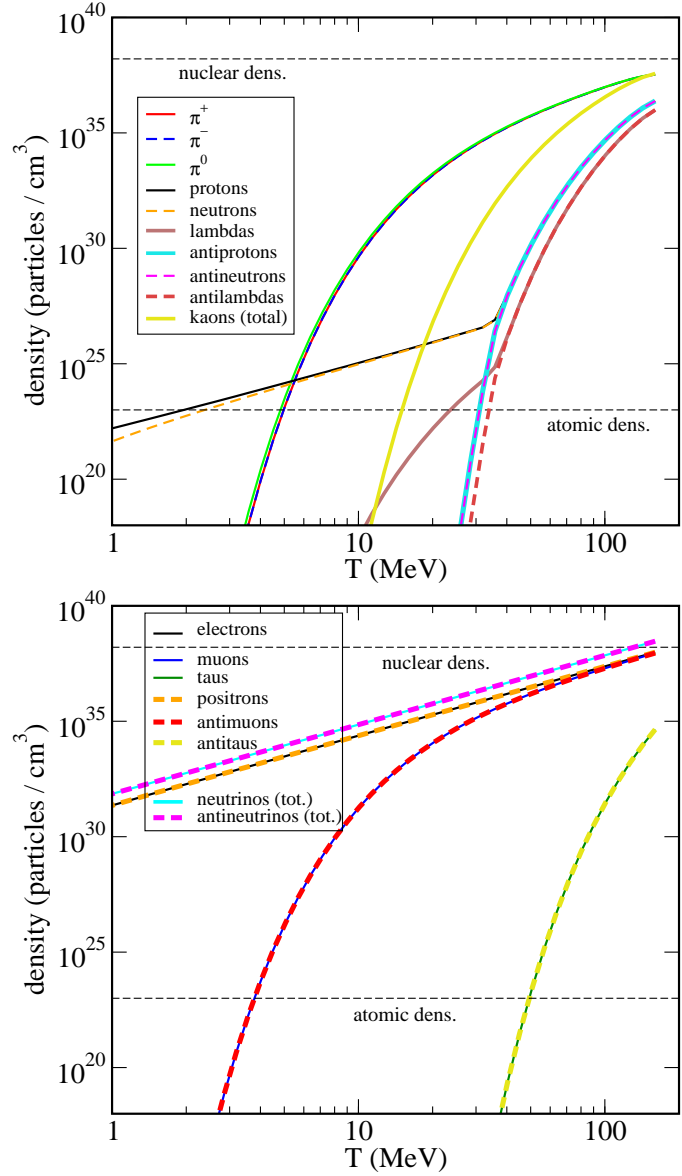

\centerline{\includegraphics[width=0.7\textwidth]{hadron_densities_APPB.eps}}
\centerline{\includegraphics[width=0.7\textwidth]{lepton_densities_APPB.eps}}
\caption{{\protect\small {(Color online) Top: The evolution of hadron densities after hadronization.  (total) indicates the sum of $K^+,K^-,K^0$ and $\overline{K}^0$.  Note pion--nucleon `equality' at $T\simeq 6$~MeV. Bottom: The evolution of lepton densities after hadronization, and (tot.) indicates the sum over neutrino flavors. }}} \label{fig:partnum}
\end{figure}

From the solution for the chemical potentials, the evolution of particle numbers follows.  These are shown in Fig.~\ref{fig:partnum}.  Strangeness, present in kaons, persists in significant number down to $T=10~{\rm MeV}$.  Pions are even more abundant, their density falling below density of nucleons only at $T\simeq 6$~MeV.  Among the leptons, note that muons also persist to relatively low temperatures in the HG phase.  Pion and muon evolution and chemical equilibration are discussed in more detail in the next Section~ \ref{sec:pionlepton}.

\begin{figure}
\centerline{\includegraphics[width=0.6\textwidth,angle=270]{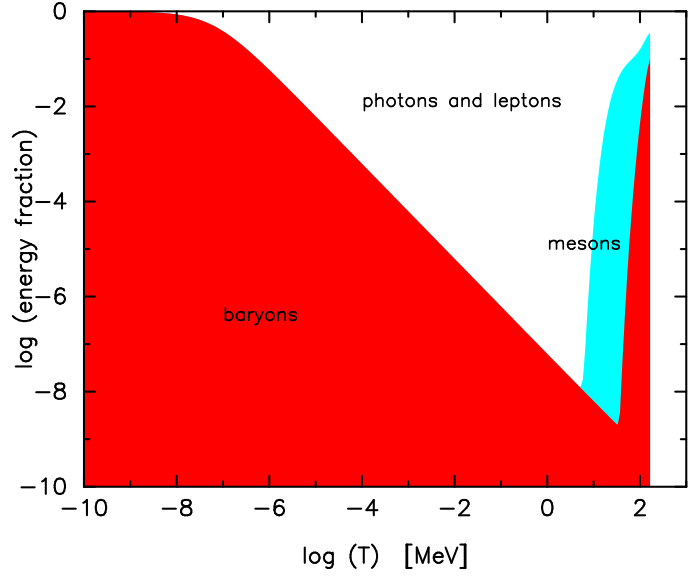}}
\caption{The fraction of the luminous energy in light particles (photons and leptons), mesons and baryons.  Note that temperature decreases from right to left.  \label{HGefrac} }
\end{figure}

Fig.~\ref{HGefrac} summarizes by showing the fraction of the luminous energy (density) in baryons as a function of temperature.  At the QGP--HG phase transformation, the fraction of energy in baryons is $\sim 10\%$, a local maximum before annihilation rapidly reduces it.  The baryons contribute significantly to the energy budget again only once the Universe has cooled and entered the matter (rest mass) dominated era.  


\subsection{Mixed phase in QGP--HG transformation}\label{MixedPhase}
It is by now well established that at low baryon density the QGP to HG phase change is a smooth transformation.  The macroscopic conditions i--iii. above are satisfied on average and for the system as a whole, but can be violated locally.

To solve the conditions Eqs.~(\ref{Q0})--(\ref{SperB}) in the simultaneous presence of both phases, the total partition function is parametrized as
\beqn
\ln{Z_{\rm tot}}=f_{\rm HG}\, 
\ln{Z_{\rm HG}}+(1 - f_{\rm HG})\, \ln{Z_{\rm QGP}} ,
\eeqn
in which $f_{\rm HG}$ represents the fraction of total phase space occupied by the HG phase.  For example, \req{Q0} is now generalized as
\begin{align}\label{Qmixed}
Q = 0 & = n_Q^{\rm QGP}\, V_{\rm QGP}+n_Q^{\rm HG}\, V_{\rm HG}, \\
  &=  V_{\rm tot} \left[ (1-f_{\rm HG})\, n_Q^{\rm QGP}+f_{\rm HG}\, n_Q^{\rm HG} \right]. \nonumber
\end{align}
and analogous expressions hold for Eqs.\,\eqref{LB0} and \eqref{SperB}.  The total volume $V_{\rm tot}$ is not relevant to the solution.

\begin{figure}[t]
\centerline{\includegraphics[width=0.58\textwidth,angle=270]{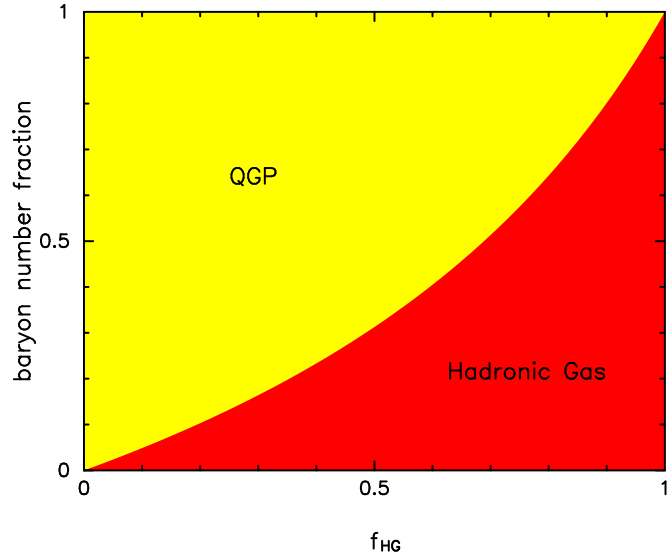}}
\caption{The fraction of baryons in each phase as a function of the parameter $f_{\rm HG}$, which is the fraction of total phase space occupied by the HG  phase.}
\label{bary_frac}
\end{figure}

Solving \req{Qmixed} and its companions determines baryon number fraction in each phase as a function of $f_{\rm HG}$, shown in Fig.~\ref{bary_frac}.  The QGP preserves a larger fraction of the baryon number density throughout the transformation, and the ratio is nearly constant $n_B^{\rm QGP} / n_B^{\rm HG} \approx 3$.

With the additional assumptions that $f_{\rm HG}$ evolves linearly in time and that the total duration of the phase transformation is $10~\mu{\rm s}$, we obtain the solution for the Universe during the transformation, indicated by the vertical lines in Fig.~\ref{chem_pot}.  In reality, both assumptions are sensitive to the equations of state of each phase and the dynamics of the phase transformation.  A more complete solution would be obtained from transport theory applied to the coexisting phases.

From this solution, the net charge per baryon $n_Q/n_B$ is calculated in each phase as a function $f_{\rm HG}$, which is independent of the additional assumptions.  Protons and neutrons being the lowest excitations in the HG phase, the HG takes on a positive charge as soon as the transformation begins.  The QGP therefore takes on a negative charge density, which is initially tiny since it occupies the larger volume, yet it can cause large variation in local electric potentials.  This charge distillation arising as a dynamical asymmetry has been discussed in the context of strangeness separation and strangelet formation~\cite{Greiner:1987tg,Greiner:1991us}

The sign of the charge distillation remains the same throughout the transformation, and therefore the total charge of the remaining QGP is increasingly negative as the transformation proceeds.  One expects the resulting electromagnetic potential to alter the chemical potentials for charged species and thereby to have a feedback effect.  Flows of charged particles will alter the uniformly small net baryon density and thus any local initial baryon--antibaryon asymmetry.  Resolving these dynamics clearly requires transport theory, as an example of the remark above.

\begin{figure}
\centerline{\includegraphics[width=0.65\textwidth]{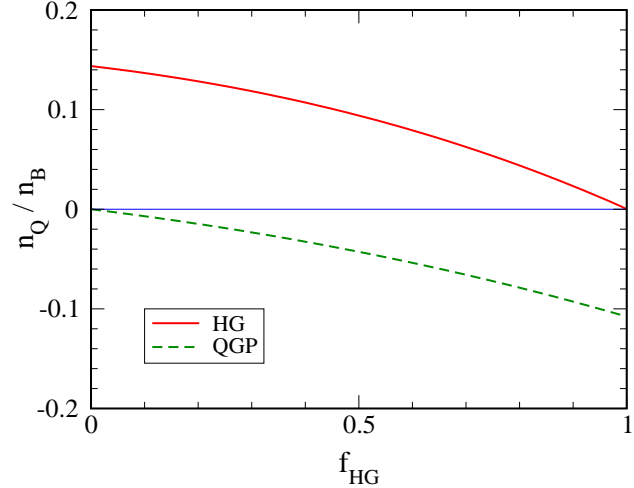}}
\caption{Net charge (including leptons) per net baryon number in the HG and QGP as function of the HG fraction $f_{\rm HG}$.  Horizontal line at zero guides the eye.}
\label{QperB}
\end{figure}

\section{Pion and lepton equilibration: $50 \gtrsim T \gtrsim 2~{\rm MeV}$}
\label{sec:pionlepton}
We now move to a closer study of the period of annihilation, decay and chemical equilibration immediately following the QGP-HG phase transformation.  As shown in Fig.~\ref{HGefrac}, this involves a large number of mesons produced during hadronization which subsequently decay, affecting the lepton populations.  The point we address here is that the same reaction that drives decay, by detailed balance, also recreates the decaying particle.  Here of great significance is that the Universe is filled with a  background gas of photons and leptons that act as a buffer to all hadron decay processes.

We demonstrate the importance of the background gas considering  the pivotal reaction 
\beqn\label{pi02gamma}
\pi^0\rightleftharpoons \gamma+\gamma,
\eeqn
which keeps hadrons in chemical equilibrium throughout the Universe evolution. This outcome is in contrast with the naive expectation that $\pi^0$ disappears, based on considering only the $\pi^0$ lifetime in vacuum $\tau_0=8.4\:10^{-17}~{\rm s}$ to the much longer time scale of the Hubble rate of expansion 
\beqn
H=1.66\sqrt{g_*}\frac{T^2}{M_{\rm Pl}}=(0.014~{\rm s})^{-1}\left(\frac{T}{\rm 10~MeV}\right)^2,
\eeqn
in which $g_*$ is the number of degrees of freedom contributing to entropy at the time (fermionic contributions are modified by the factor 7/8) and $M_{\rm Pl}=1.22\:10^{19}~{\rm GeV}$ is the Planck mass.

Note that the large pion abundance that goes along with  reaction \req{pi02gamma}  can influence the relic neutrino density and potentially even BBN  which immediately follows this `pion' era -- we hope to return to these topics in the near future. Reaction \req{pi02gamma} is example of important family of one-to-two  reactions which have only be studied recently~\cite{Kuznetsova:2008jt,Kuznetsova:2010pi}.  Another example is the decay of charged pions $\pi^{\pm}$ can impact the residual populations of muons, electrons and neutrinos both through decay and scattering
\begin{align}\label{chargedpidecay}
\pi^{\pm} \rightleftharpoons \mu^{\pm}+\nu_\mu(\bar\nu_\mu).
\end{align}
The decay of muons also contributes to neutrino density
\beqn\label{mudecay}
\mu^{\pm}\rightleftharpoons e^{\pm}+\nu_e(\bar\nu_e)+\bar\nu_\mu(\nu_\mu).
\eeqn

Two-to-two reactions maintain equilibrium within and between charged pion populations
\beqn\label{chargedpipi0}
\pi^0+\pi^0 \rightleftharpoons \pi^++\pi^-, \quad 
\gamma+\gamma \rightleftharpoons \pi^++\pi^-,
\eeqn
and lepton populations
\beqn\label{2lto2l}
\begin{split}
e^++e^-\rightleftharpoons \mu^++\mu^-,
\quad \gamma+\gamma\rightleftharpoons l^++l^-,\\
\pi^++\pi^-\rightleftharpoons l^++l^-,~~(l=\mu,e).
\end{split}
\eeqn
In the following we describe the chemical relaxation times of each of these reactions and compare it to the other relevant timescales, especially the expansion of the Universe.   

Consider the first example, \req{pi02gamma} controlling $\pi^0$ chemical equilibration with the background of thermal photons.  The population equation describing the evolution of $\pi^0$ number can be cast into the form a differential equation for the $\pi^0$ fugacity $\Upsilon_{3}$:
\beqn\label{upsilondot}
\frac{d}{dt}\Upsilon_{3}=\frac{1}{\tau_T}\Upsilon_3 +\frac{1}{\tau_S}\Upsilon_3
  + \frac{1}{\tau_3}(\Upsilon_1\Upsilon_2 - \Upsilon_{3}),
\eeqn
see discussion in~\cite{Kuznetsova:2008zr}, partially recounted in the appendix.
We keep numerical subscripts here rather than simplifying with $1=2=\gamma$ so that later generalization is clear.  Here, the $\tau_T$ and $\tau_S$ are kinematic relaxation times for the temperature and entropy:
\begin{align}
\frac{1}{\tau_T}\:&=-g_*T^3\left(\frac{dn_3}{d\Upsilon_3}\right)^{\!-1}\frac{d}{dT}\left(\frac{n_3}{\Upsilon_3g_*T^3}\right)\frac{dT}{dt} \label{tauT} ,\\
\frac{1}{\tau_S}\:&=-\frac{n_3}{\Upsilon_3}\left(\frac{dn_3}{d\Upsilon_3}\right)^{\!-1}\frac{d\ln (g_*T^3V)}{dT}\frac{dT}{dt} \label{tauS}.
\end{align}
The leading minus signs mean that $\tau_T,\tau_S>0$.  

$\tau_3$ is the chemical relaxation time incorporating effects of the background gas~\cite{Kuznetsova:2010pi}:
\beqn\label{tauidefn}
\tau_3=\frac{1}{V}\frac{dN_3}{d\Upsilon_3}\Upsilon_3\left(\frac{dW_{3\rightarrow 12}}{dVdt}\right)^{-1}=\frac{1}{V}\frac{dN_3}{d\Upsilon_3}\Upsilon_1\Upsilon_2\left(\frac{dW_{12\rightarrow 3}}{dVdt}\right)^{-1},
\eeqn
where $N_3$ is pion number.  $dW/dVdt$ is the rate per unit volume for the process in subscript accounting for the presence of the background gas of reactants and products, and which therefore contains dependence on the corresponding fugacities of the reaction participants.  Thus, in the Boltzmann limit the combination $\Upsilon_3(dW_{3\to 12}/dVdt)^{-1}$ is independent of the fugacity.  The second equality follows from the detailed balance condition, \req{detailedbal}. 

Entropy is conserved in the expanding Universe, and in the radiation dominated Universe we have 
$d(T^3V)/dT =0$, and hence $1/\tau_S=0$ for the present discussion.  $1/\tau_T$ describes the effect of dilution of phase space and can be dominant in situations where the number of  the type `3' particles is preserved but their density decreases due to expansion of the volume.  In the early Universe, comparing $\tau_T$ to the chemical relaxation time $\tau_3$  provides the quantitative condition for freeze-out from chemical equilibrium 
\beqn\label{freezeout}
\tau_3\simeq \tau_T\simeq \frac{T}{m_3H}.
\eeqn
In the second equality, we have given the relevant estimate of $\tau_T$ for the early Universe~\cite{Kuznetsova:2010pi}.

\begin{figure}[tbp]
\centerline{
\includegraphics[width=0.50\textwidth]{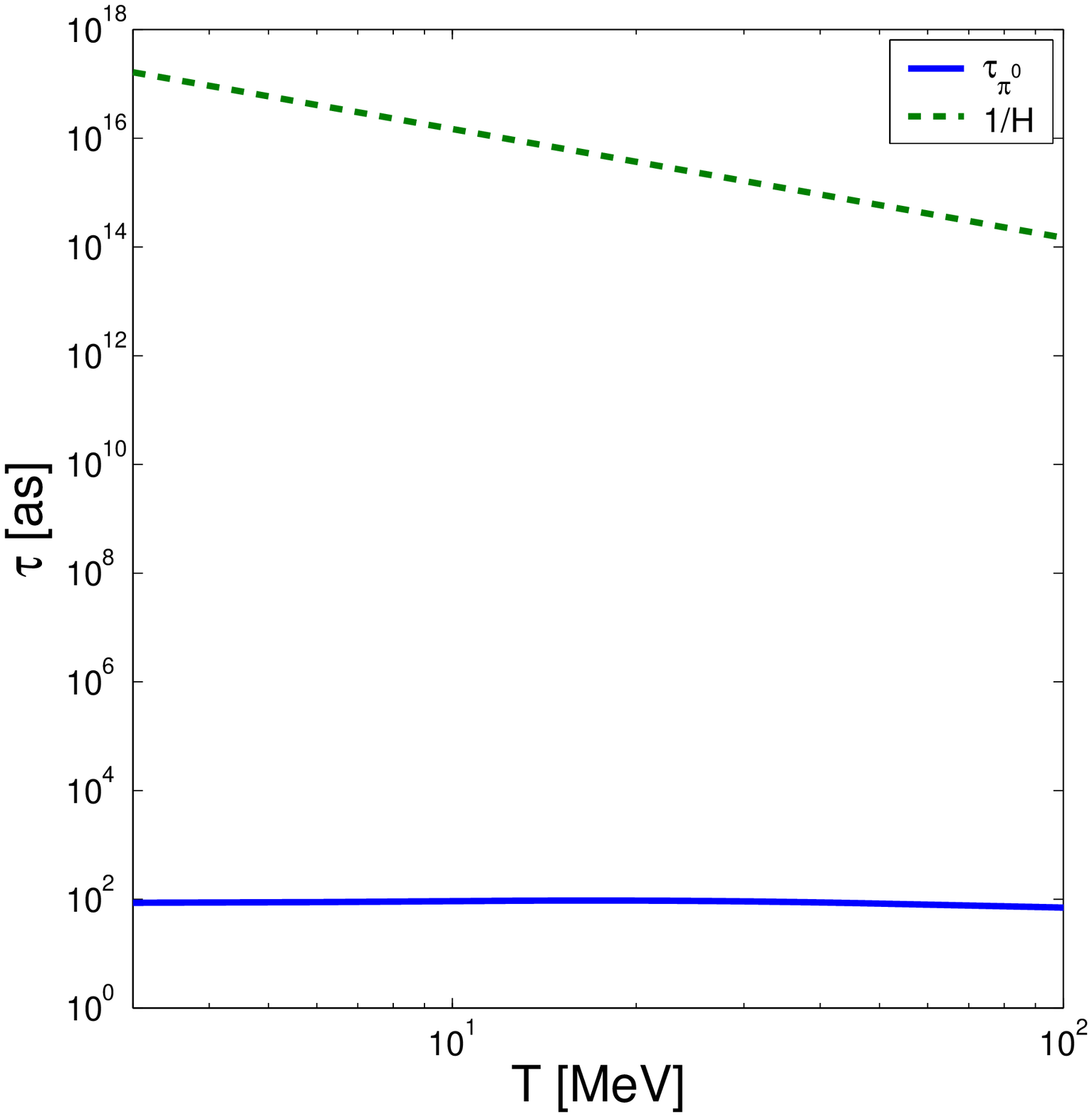}\hspace*{-0.3cm}
\includegraphics[width=0.46\textwidth]{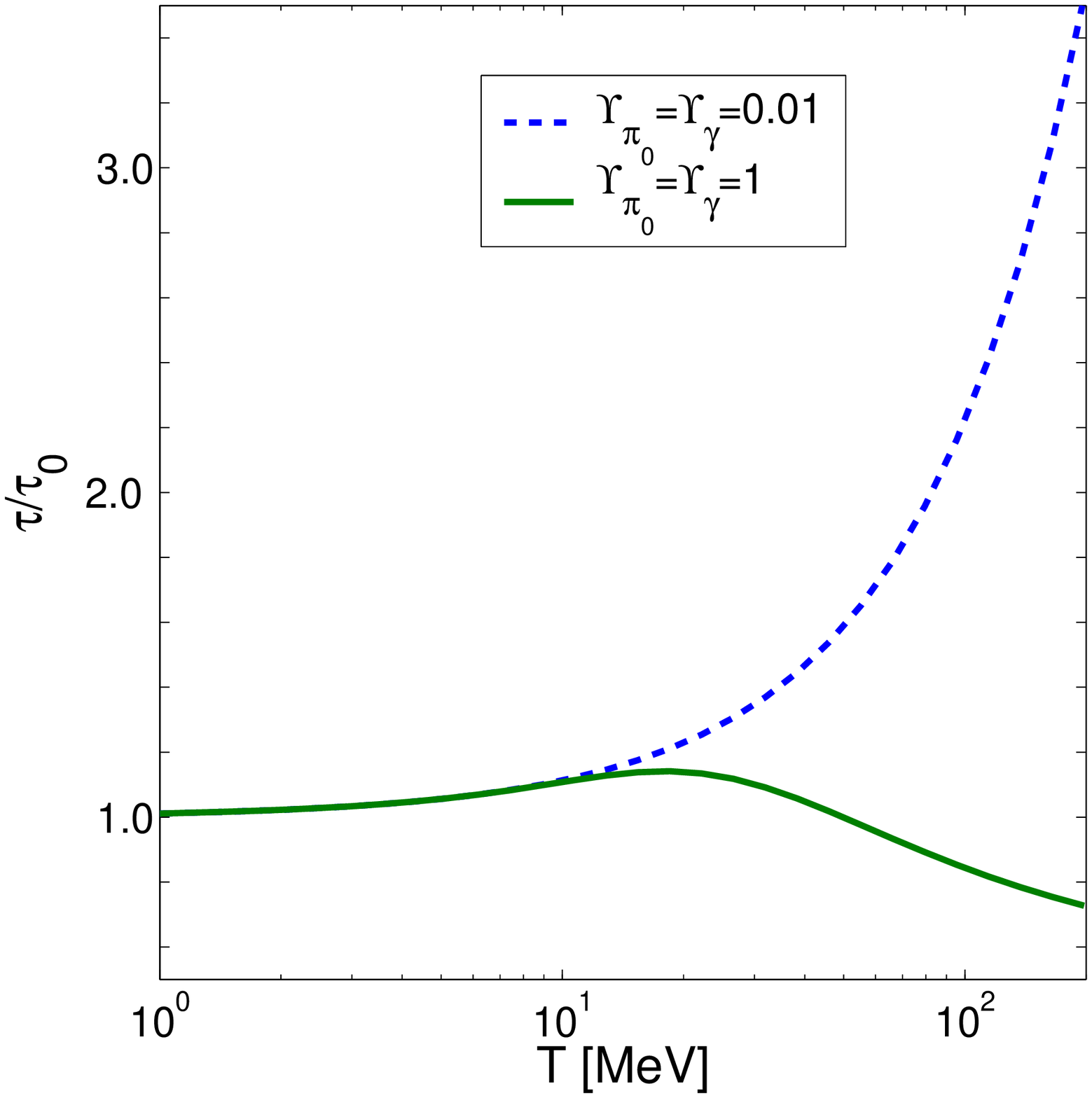}
}
\caption{{\protect\small {(Color online) At left: The Boltzmann (low phase space occupancy) limit [dashed (blue) line] to the quantum statistical gas [solid (green) line] with fugacity $\Upsilon_f=1$.  At right: $\pi^0$ equilibration time $\tau_3$ [solid (blue) line] and Universe expansion time $1/H$ [dashed (green) line] as functions of temperature.  }}} \label{fig:pi0}
\end{figure}

We can now study the reaction \req{pi02gamma} quantitatively.  In the early Universe, the number of photons is high and is characterized by the observed value of $\eta$, see below~\req{SperBcalc}, and pions are also copious immediately after the phase transformation. These conditions, specifically chemical equilibrium of the HG, mean the fugacities $\Upsilon_f$ are unity and therefore the quantum statistics of the gases are important, leading to bose enhancement and fermi blocking as the case may be, recognized by Uehling and Uhlenbeck~\cite{BUU}.  The importance of this effect is shown at left in Figure~\ref{fig:pi0}.  

Comparing $\tau_3$ to $\tau_T$ ($1/H$) at right shows that $\pi^0$ remains in chemical equilibrium even as its thermal number density gradually decreases, consistently with  falling thermal production rates.  Even so at all times $\pi_0$ remain in chemical equilibrium: this phenomenon can be attributed to the high population of photons, within which it remains probable to find photons of high enough energy to fuse into $\pi^0$, and as the number of high energy photons decreases, described by photons' Planck distribution, so does the number of $\pi_0$ which need to be maintained.  

One can generalize \req{upsilondot} for the three-body muon decay \req{mudecay} and the 2-to-2 reactions Eqs.\:\eqref{chargedpipi0},\eqref{2lto2l} relevant for pion and lepton equilibration~\cite{Kuznetsova:2008zr,Kuznetsova:2010pi,Kuznetsova:2008jt}.  For example, for the three-body decay there would be three numerical subscripts $1,2,3$ on the right, each for a distinguishable particle species and the product of their three fugacities appears in the equation for chemical relaxation time, $\tau_4$, similar to Eq.(\ref{tauidefn}).

From this study, we learn several important characteristics of the plasma present in the Universe preceding and up to the time of BBN.  First, $\pi^{\pm}$ and $\mu^{\pm}$ remain in chemical equilibrium until $T\simeq 6~{\rm MeV}$ and $\sim$4~MeV, respectively~\cite{Kuznetsova:2010pi}, alone due to the specific reactions considered -- if and when other reactions are important the actual freeze-out temperature could be still lower.  Note here that we evaluate the reaction rate for muons \req{mudecay} at low temperatures $T\ll m_\mu$ based on the muon lifespan in vacuum $2.20\:10^{-6}~{\rm s}$,  muon equilibrium is therefore preserved due to detailed balance. However, muon decay is a process in which helicity considerations keep the reaction slow, and in the dense early Universe plasma muon decay rate can increase, which by detailed balance argument implies that the back reaction also is enhances, and the chemical equilibrium is maintained longer.  

All this means that  a significant number of $\mu^{\pm}$ is  present along with $\pi^0$ to interact with nucleons even at low (but not necessarily BBN-range ) temperatures, as shown at left in Figure~\ref{fig:partnum}.  A further consequence of the maintenance of $\pi,\mu$ chemical equilibrium is a relatively large effective degeneracy, implied also by the significant fraction of mesons in the energy budget, Fig.~\ref{HGefrac}.

\begin{figure}[tbp]
\centerline{
\includegraphics[width=0.65\textwidth]{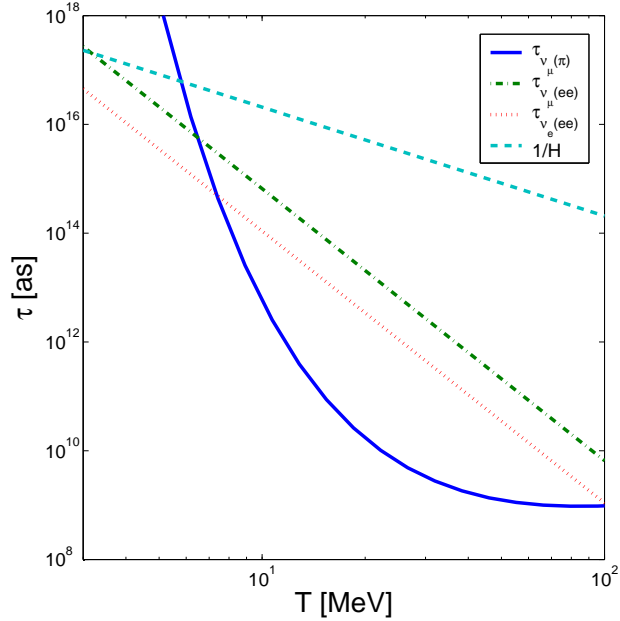}
}
\caption{{\protect\small {(Color online) 
Neutrino reaction relaxation times compared to Hubble rate.  Solid (blue) line is for \req{chargedpidecay} while dotted (red) and dash-dotted (green) are for the electron-- and muon--neutrino processes \req{ee2nunu}.  }}} \label{fig:numupi}
\end{figure}

From neutrino population dynamics, we learn that relatively heavy particles ($m\gg T$), in this case pions and muons, can be important influences on light particles' equilibration and population.  At right, in Figure~\ref{fig:numupi}, one can see how important the pion-muon-neutrino reaction \req{chargedpidecay} is in maintaining neutrino chemical equilibrium in this epoch.  It appears to be more important than
\beqn\label{ee2nunu}
e^++e^-\rightleftharpoons \nu_{e,\mu}+\bar\nu_{e,\mu},
\eeqn
whose rate is obtained in the limit of small neutrino chemical potential $\mu_\nu\ll T$~\cite{Freese:1982ci} (shown as the dotted and dashed dotted lines).

We have only highlighted here a few outcomes of the study to illustrate the necessity of studying quantitatively the chemical evolution and equilibration of the early Universe plasma.  More results and discussion are found in~\cite{Kuznetsova:2010pi}.

\section{Conclusions}\label{sec:concl}
We have shown that hadrons remain in chemical equilibrium throughout the evolution of the Universe. The same mechanism that allows $\pi_0$ to rapidly decay in reaction \req{pi02gamma} is  present when two photons collide in the thermally equilibrated Universe, and photon `fusion' reactions fill any missing $\pi_0$.  In order to see that there is no freeze-out of pions,   a  quantitative study is required~\cite{Kuznetsova:2008jt,Kuznetsova:2010pi}.

Given that hadron chemical equilibrium is maintained, one can use methods of hadro-chemistry to study particle abundances. We have shown how to compute the values of the quark and lepton chemical potentials that yield the observed matter--antimatter asymmetry in the Early Universe after hadronization and equilibration of the HG.  Furthermore, we have seen that the non-zero chemical potentials drive charge distillation during the phase transformation, with the QGP and HG having negative and positive charge densities, respectively.  Separation of baryons and antibaryons into domains could maintain a homogeneous zero charge density Universe, a phenomenon which could, {e.g.}, play a significant role in amplifying a pre-existent, much smaller net baryon yield.

We have described the chemistry of the HG in the Universe evolution period down to $T\sim 3~{\rm MeV}$ noting a significant yield of pions and muons which remain in chemical equilibrium throughout.  Consequently, they remain a significant fraction of the particle number and energy in pre-BBN Universe. Their presence and reactions  can impact the relic neutrino population.

The lessons of this study are manifold, perhaps the most salient of which is the importance of studying the chemistry of the early Universe quark and hadron plasma quantitatively.  Future work will be necessary to explore the implications of the findings here for, e.g., baryon separation, relic neutrino background,  and thus big bang nucleosynthesis.

\begin{appendix}
\section{Fugacity Evolution Equation}
Consider the case of an unstable particle 3 with a two-body decay to particles 1 and 2.  The number of `3' particles is governed by the population equation
\beqn\label{1to2popeq}
\frac{1}{V}\frac{dN_{3}}{dt}=
    \frac{dW_{12\rightarrow 3}}{dVdt}-\frac{dW_{3\rightarrow 12}}{dVdt},  
\eeqn
which gives the change in number of particle 3 as increased by production (particles 1 and 2 fuse into 3) and decreased by the two-body decay $3\to 1+2$. 

In the absence of ambient populations of particles 1, 2, 3, there will be no production and decay will occur as in vacuum.  In this case, \req{1to2popeq} reduces to $V^{-1}dN_3/dt\simeq -(\gamma\tau_0)^{-1} N_3/V$, where $\tau_0$ is the lifetime of particle 3 in vacuum, and results in the usual exponential decay of particle 3 number.  Here, $\gamma$ is the Lorentz boost factor relating the particle rest frame (where $\tau_0$ is determined) to the observer frame; in our case, the observer is the heat bath of the Universe.  

The explicit forms of the rates $\frac{dW_{12\rightarrow 3}}{dVdt}$ and $\frac{dW_{3\rightarrow 12}}{dVdt}$ are discussed in~\cite{Kuznetsova:2008zr,Kuznetsova:2010pi}.  Considering their structure, energy--momentum conservation and time-reversal symmetry of the $3\rightleftharpoons 1+2$ matrix element together imply the detailed balance relation
\beqn\label{detailedbal}
\frac{dW_{12\rightarrow 3}}{dVdt}\Upsilon_{3}=
\frac{dW_{3\rightarrow 12}}{dVdt} \Upsilon_{1}\Upsilon_{2}.
\eeqn
Therefore, in chemical equilibrium, when $dN_3/dt=0$, \req{1to2popeq} shows that the fugacities must satisfy
\beqn\label{equilcond}
\Upsilon_{3}=\Upsilon_{1}\Upsilon_{2},
\eeqn
which corresponds to the Gibbs condition for the chemical potentials.

Using \req{detailedbal}, one can rewrite \req{1to2popeq} into the differential equation for $\Upsilon$ presented in \req{upsilondot}.
 
\end{appendix}
\vskip4mm
\noindent {\it Acknowledgments} 
This work has been supported by a grant from the U.S. Department of Energy, DE-FG02-04ER41318; Laboratoire de Physique Th{\' e}o\-rique et Hautes Energies, LPTHE, at University Paris 6 is supported by CNRS as Unit{\' e} Mixte de Recherche, UMR7589. 


\end{document}